\documentstyle{lamuphys}
\input{epsf.sty}
\newcommand{\qbar}{{\overline{q}}}
\newcommand{\qt}{{\tilde{q}}}
\newcommand{\qtbar}{{\overline{\tilde{q}}}}
\newcommand{\Dsl}{/\!\!\!\! D}
\newcommand{\etat}{{\tilde{\eta}}}
\newcommand{\half}{\frac{1}{2}}

\makeatletter
\let\chapter\hid@chapter
\makeatother
\begin{document}
\pagenumbering{arabic}
\title{Connections Between Lattice Gauge Theory and Chiral Perturbation Theory}

\author{Maarten\,Golterman}

\institute{Department of Physics, Washington University,
St. Louis, MO 63130, USA\\
e-mail: maarten@aapje.wustl.edu}

\titlerunning{Lattice Gauge Theory and ChPT}
\maketitle

\begin{abstract}
In this talk, I address the comparison between results from
lattice QCD computations and Chiral Perturbation Theory (ChPT).
I briefly discuss how ChPT can be adapted to the much-used
quenched approximation and what it tells us about the
special role of the $\eta'$ in the quenched theory.  I then
review lattice results for some quantities (the pion mass,
pion scattering lengths and the $K^+\to\pi^+\pi^0$ matrix
element) and what quenched ChPT has to say about them.
\end{abstract}
\section{Introduction}

Chiral Perturbation Theory (ChPT) (\cite{WeS79,GaLe85,Ga97}) gives us 
information about the functional dependence of quantities 
associated with low-energy Goldstone Boson (GB) physics
 on the light-quark masses. Examples
of such quantities are the GB masses, decay constants and scattering
amplitudes.  At any given order, these relations between physical
quantities and quark masses involve a finite number of constants (the
``low-energy constants (LECs)," which cannot be determined from
ChPT alone.  Therefore, ChPT is predictive (at any given order) if
we consider a number of physical quantities larger than the number
of LECs needed at that order. 
 These LECs can in principle be determined either by
comparison with experimental data, or by a theoretical calculation from
the underlying theory, the Standard Model.

The strong-interaction part of such calculations is nonperturbative,
and this is where Lattice QCD (LQCD) comes in.  In LQCD, physical
quantities (or related quantities, such as weak matrix elements) are
computed from first principles as a function of the quark masses.
By fitting the results with the relations predicted by ChPT, one can
then, in principle, determine the LECs. This is very similar to 
determining these constants from experimental data, with the added
advantage that in LQCD one can vary the quark masses.

Of course, in general, LQCD results will need to be of a high precision 
in order to extract the $O(p^4)$ LECs, because, in general, they show up in  
the one-loop corrections to the tree-level predictions from
ChPT.  This means good control over both statistical and systematic
errors in the lattice computations.  For instance, we need the 
volume $L^3$ in lattice units to be large
in order to use a small lattice spacing $a$, while keeping the
physical volume large enough to fit the hadronic system of interest.
It is important to keep in mind that lattice results need to be
extrapolated to the continuum limit before they can be compared with
ChPT.  Also, for ChPT to be valid, GB masses need to be small compared
to the chiral symmetry breaking scale.  Again, this leads to the
requirement of large enough volumes.  (Finite volume effects
can be studied within ChPT (\cite{Ga87}); we will see some
examples in the following.  However, LQCD practitioners are not only
interested in the comparison with ChPT!) The temporal extent of the
lattice needs to be large enough so that the lowest state in any
channel can be reliably projected out by taking the large-time limit
of euclidean correlation functions.

This has led to the widespread use of the {\it Quenched Approximation}
(QA) (\cite{Pa81,WeD82}), 
in which the quark determinant is omitted from the LQCD
path-integral.  This is equivalent to omitting all contributions
to correlation functions that involve sea-quark loops (valence-quark 
loops resulting from contractions of quarks in composite operators
of course remain present, as they have nothing to do with the
determinant).  The reason is that, if the fermion determinant is
included, the necessary computer time increases by orders of 
magnitude if we keep the physical parameters (lattice spacing,
volume, quark masses) the same.

Quenched QCD is a different theory from QCD (as we will see, it's not
even a healthy theory!), and therefore the predictions of ChPT do
not apply to lattice results obtained in the QA.  Fortunately, it
turns out that a quenched version of ChPT (QChPT) can be developed
systematically, and it is in this framework that we can compare
quenched LQCD with ChPT. Unfortunately, this also implies that the
LECs predicted by quenched QCD are not necessarily equal to those
of full QCD, and as such, knowledge of them is of somewhat limited
value. 
In this talk, I will describe how QChPT 
works, and discuss some examples of the comparison of lattice
results with ChPT.  I should note right away
that, as we will see, currently
lattice data are not precise enough yet to 
unambiguously see ChPT one-loop effects. 

Before we get into this, let me end this introduction with a few
remarks.  First, recently, more lattice results with ``dynamical
fermions" (i.e. including sea-quark loops) are becoming available.
However, since the overhead in computing the fermion determinant
is so large, these results are often for one or two values of
the sea-quark mass, while many values of the valence quark masses
are considered.  The methods described in this talk can easily
be adapted to these ``partially quenched" theories with sea
quarks with a mass that differs from that of the valence quarks
(\cite{BeGo94,Sh972}).  
Second, the study of QChPT sheds much light on the nature
of the QA, and, as such, has been very helpful for LQCD.

\section{Quenched ChPT, the $\eta'$, and the Pion Mass}
Euclidean quenched QCD can be defined by taking the ordinary
QCD lagrangian, and adding a new set of quarks 
$\{\qt_i\}$ to it which carry
one-by-one exactly the same quantum numbers as the normal quarks
$\{q_i\}$, but which have opposite statistics (\cite{Mo87}):  
\begin{equation}
{\cal L}=\qbar_i(\Dsl+m_i)q_i+\qtbar_i(\Dsl+m_i)\qt_i\;,
\ \ \ i=u,d,s.  \label{lag}
\end{equation}
We will refer
to these new wrong-statistics quarks as ``ghost-quarks." 
The reason for their introduction is that 
the contribution coming from integrating over these 
ghost-quarks exactly cancels the determinant which comes
from the integration over the normal quarks.  It is easy to
see this diagrammatically: while a minus sign is needed for
every occurence of a quark loop, this is not the case for
a ghost-quark loop, because of their bosonic statistics.
So, for every diagram with a quark loop, a similar diagram
contributes with that one loop replaced by a ghost-quark
loop, therefore leading to the cancellation of this diagram.

For $m_i=0$, the lagrangian (\ref{lag}) is invariant under
a {\em six}-flavor chiral symmetry group (\cite{BeGo92}).
A typical transformation looks like
\begin{equation}
\pmatrix{q'_i\cr\qt'_i}=\pmatrix{A_{ij}&B_{ij}\cr
C_{ij}&D_{ij}\cr}\pmatrix{q_j\cr\qt_j}, \label{transf}
\end{equation}
where $A$, $B$, $C$ and $D$ are $3\times 3$ blocks, with
$B$ and $C$ containing anticommuting numbers, since they
transform bosons into fermions and vice versa.  This group
is known as a {\em graded} version of U(6), and is denoted
by U(3$|$3).  For $m_i=0$, the symmetry of (\ref{lag}) is
therefore U(3$|$3)$_L\times$U(3$|$3)$_R$.

We will assume that gluons are responsible for forming
mesonic bound states.  In this case, since gluons couple
equally to quarks and ghost-quarks, that means that not only
the usual $q\qbar$ GBs (i.e. $\pi$, $K$, $\eta$ and
$\eta'$) will occur, but also $\qt\qtbar$ ``ghost-mesons,"
and $q\qtbar$, $\qt\qbar$ {\em fermionic} ``hybrid" mesons.
(For the role of $\eta'$, see below.)  QChPT can then be
developed systematically in a way completely analogous
to the usual three-flavor case (\cite{BeGo92}, for other
early work on QChPT, see Sharpe (1990,1992)), 
but now for the group
U(3$|$3)$_L\times$U(3$|$3)$_R$.  The quenched effective
lagrangian describes the low-energy physics of all
Goldstone mesons, ghosts and fermionic hybrids included.

Lack of space prevents me from giving more details on the
technicalities of QChPT (see e.g. \cite{Go94}).  However, 
before we continue to consider any examples, we need to 
address the very special role of the $\eta'$.  In unquenched
QCD, the $\eta'$ is not a GB because of the U(1)$_A$ anomaly.
In quenched QCD, we would expect that a similar role is
played by some linear combination of $\eta'$ and $\etat'$.
We can understand which linear combination by realizing that
ghost-quark loops do {\em not} contribute a minus sign, so that
we get a nonzero triangle anomaly between the U(1) axial 
current and two gluons by {\em subtracting} the triangle
graphs with a ghost-quark running around the loop from those
with normal quarks.  Hence in the quenched theory, the 
``anomalous" meson is the ``super-$\eta'$", $(\eta'-\etat')/
\sqrt{2}$. The super-$\eta'$ transforms as a singlet under
the nonanomalous part of the graded chiral symmetry group, and
therefore arbitrary functions of this field (and its
derivatives) appear 
in the effective lagrangian (\cite{GaLe85,BeGo92}).

Let us consider the lowest order, $O(p^2)$ part of the
effective lagrangian quadratic in the $\eta'$ and $\etat'$
fields:
\begin{eqnarray}
{\cal L_{\eta'}}&=&\half(\partial_\mu\eta')^2-
\half(\partial_\mu\etat')^2+\half m_\pi^2(\eta'^2-
\etat'^2) \label{letap} \\
&&+\half\mu^2(\eta'-\etat')^2+\half\alpha
(\partial_\mu(\eta'-\etat'))^2, \nonumber
\end{eqnarray} 
where the terms on the second line are the quadratic,
$O(p^0)$ and $O(p^2)$ parts of the arbitrary functions 
of the super-$\eta'$ field.  All minus signs originate
from the graded nature of the chiral symmetry group
(in building invariants for such a group, the supertrace
plays the role of the trace in the ordinary case, see
\cite{DeW}). 
They will have important consequences.  For simplicity
we assumed degenerate quark masses.

If we omit the $\etat'$ field from (\ref{letap}),
we obtain the quadratic part of ${\cal L_{\eta'}}$
for the unquenched theory, and the parameter $\mu^2$ would
essentially be the singlet part of the $\eta'$ mass
(one can show that $\mu^2$ does not vanish because of
the anomaly!).  So, let us see what happens in the
quenched case.  We will treat the first line in (\ref{letap})
as the part defining $\eta'$ and $\etat'$ propagators
$S^0_{\eta'}(p)$ and $S^0_{\etat'}(p)$:
\begin{equation}
S^0_{\eta'}(p)=-S^0_{\etat'}(p)=\frac{1}{p^2+m_\pi^2}\;. 
\label{etaprop}
\end{equation}
The second line defines the two-point vertices 
$-(\mu^2+\alpha p^2)$ for $\eta'$-$\eta'$, $\etat'$-$\etat'$
and for $\eta'$-$\etat'$ mixing.  We can find the complete $\eta'$
two-point function by summing all diagrams containing an arbitary
number of these two-point vertices on an $\eta'$ line
(taking into account combinatoric factors):
\begin{eqnarray}
S_{\eta'}(p)&=&S^0_{\eta'}(p) \label{etapropsum} \\
&&-S^0_{\eta'}(p)(\mu^2+\alpha p^2)S^0_{\eta'}(p) \nonumber \\
&&+S^0_{\eta'}(p)(\mu^2+\alpha p^2)(S^0_{\eta'}(p)+
S^0_{\etat'}(p))
(\mu^2+\alpha p^2)S^0_{\eta'}(p) \nonumber \\
&&-\cdots. \nonumber
\end{eqnarray}
Using (\ref{etaprop}), we see that the third line in
(\ref{etapropsum}) vanishes (and so do all higher order 
contributions), leading to
\begin{equation}
S_{\eta'}(p)=\frac{1}{p^2+m_\pi^2}-\frac{\mu^2+\alpha p^2}
{(p^2+m_\pi^2)^2}. \label{etapropfull}
\end{equation}
It is actually easy to convince oneself, that this cancellation
is nothing else than the cancellation of sea-quark loops with
their ghost counterparts.  The vertex $-(\mu^2+\alpha p^2)$
represents the annihilation of the valence quark antiquark pair
in the $\eta'$ (which is a flavor singlet, making this
annihilation possible), and the creation of either a 
quark antiquark, or a ghost-quark ghost-antiquark pair.  
Tracing the quark flow for each of the terms in (\ref{etapropsum}),
one finds that quark and ghost-quark loops occur for all terms
with more than one vertex insertion.

 From (\ref{etapropfull}), it is clear that the $\eta'$ is ``sick"
in the QA because there is no particle interpretation for the
double pole term.  Moreover, it tells us that we
{\em have to keep} the $\eta'$ in the QA, because it has poles
which are degenerate with those of the other Goldstone mesons.
These poles can lead to chiral logarithms that have no
counterpart in the unquenched theory. In the case of
nondegenerate quark masses, the $\pi^0$ and $\eta$ will
inherit the problems of the $\eta'$ through mixing.

As an example, consider the one-loop expression for the
pion mass in terms of the quark mass, calculated in QChPT:
\begin{eqnarray}
m_\pi^2&=&Am_{\rm q}\left(1+\delta\log{Bm_{\rm q}}+
Cm_{\rm q}\right), \label{pionmass} \\
\delta&=&\frac{\mu^2}{24\pi^2 f_\pi^2}, \label{delta}
\end{eqnarray}
where, for simplicity, we set $\alpha=0$ (of course in
comparing with real data, one has to take $\alpha$ into
account).  $A$, $B$ and $C$ are parameters related to the
LECs, which are not predicted by ChPT. We see that the
chiral log in (\ref{pionmass}) is entirely due to the
double pole term in the $\eta'$ propagator (because it is
proportional to $\mu^2$).  In unquenched ChPT, the chiral
log would be multiplied by an extra factor $m_{\rm q}$,
and be more suppressed for small quark mass.  In fact,
in the quenched case,
the quantity $m_\pi^2/m_{\rm q}$ diverges in the chiral limit,
$m_{\rm q}\to 0$! This is the first example of the highly
infrared divergent behavior of quenched QCD. (I believe that
this divergence is a sickness of quenched QCD, and not just
of QChPT (\cite{BeGo93}).)

\begin{figure}[t]
\vspace{-2.5cm}
\epsfxsize=1.0 \hsize
\epsfbox{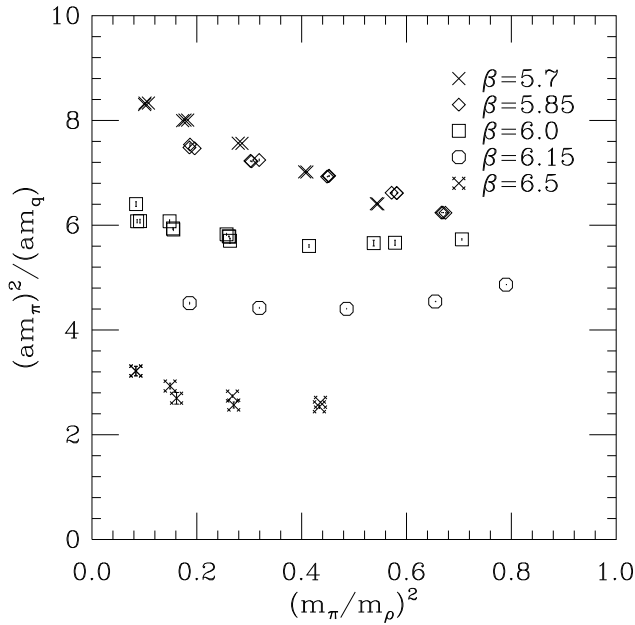}
\caption[]{Quenched staggered fermion lattice results for 
$m_\pi^2/m_{\rm q}$, in lattice units, versus the quark
mass in dimensionless units (see text).  From
the Lattice 96 spectrum review (\cite{GoS97}).}
\end{figure}

In Fig. 1, I show lattice results for this quantity (in 
lattice units), from a compilation of staggered fermion
spectrum results by \cite{GoS97} (see this review for the
original refs. on all these data).  The mass ratio on the
horizontal axis is basically the quark mass.  $\beta=5.7$
corresponds to a lattice spacing $a\approx 0.2$~fm, while
$\beta=6.5$ corresponds to $a\approx 0.055$~fm. The squares
and fancy crosses correspond to the data with the largest
spatial volumes.  The overlapping squares have 
$L\approx 2.6$~fm and $L\approx 3.5$~fm, while the highest
square (for lowest quark mass) has $L\approx 1.8$~fm.  The
fancy crosses have $L\approx 2.6$~fm.  The fact that the
squares for $L\approx 2.6$~fm and $L\approx 3.5$~fm overlap
indicates that $L\approx 2.6$~fm is large enough to not have
finite volume effects larger than the size of the error bars.
Points for different $\beta$ do not fall onto
one curve because the quantity on the vertical axis depends
on $a$.  While the data do show an upward trend with decreasing
quark mass, as predicted by (\ref{pionmass}), we cannot
conclude that this behavior is unambiguously seen in these data.
We know that scaling violations occur for the data at the two
lower $\beta$ values, so that these cannot be reliably compared
to the continuum prediction of QChPT.  At larger $\beta$, the
curves look flatter, and more data for small quark masses at 
these higher values of $\beta$ will be needed to fit 
(\ref{pionmass}). Qualitatively, we do see a departure from
the ChPT tree-level prediction $m_\pi^2\propto m_{\rm q}$, which
would correspond to a horizontal line in Fig. 1.  For a recent,
much more detailed discussion of attempts to fit (\ref{pionmass})
to lattice data, including a list of subtleties and pitfalls,
see \cite{Sh971}.  We note here that it is probably better to
determine $\delta$ from a certain ratio of decay constants
(\cite{BeGo93,Ba96,Sh971}). 

The logarithm in (\ref{pionmass}) appears with the coefficient
$\delta$, instead of the usual $m_\pi^2/(4\pi f_\pi)^2$, and 
therefore is of the same order as the tree-level term in chiral
power counting.  At more than one loop, there are contributions
of the same order in $m_\pi^2/(4\pi f_\pi)^2$, but higher order
in $\delta$.  This means that, in order for QChPT to be systematic,
$\delta$ has to be treated as a small parameter.  In unquenched
QCD, we can estimate $\delta$ from the $\eta'$ mass, yielding
$\delta\approx 0.18$.  Of course, since the quenched theory is
different, $\delta$ could have a different value in this case,
and it is important to determine its value numerically.  
It is believed that the quenched value does not differ much
from the unquenched value, but errors are not yet sufficiently
under control to quote a number (\cite{Sh971}).   We also note
here that both $\mu^2$ and $\alpha$ are of order $1/N$ in the
large-$N$ expansion (\cite{Ve79,Wi79}).

\begin{figure}[t]
\epsfxsize=0.8 \hsize
\epsfbox{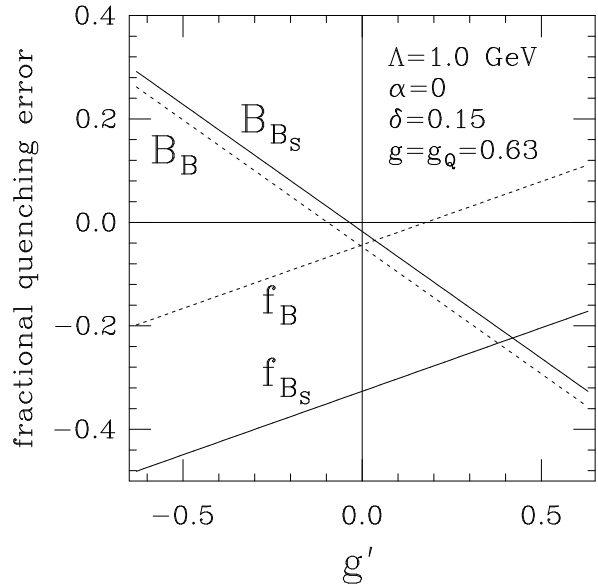}
\vspace{-1.0cm}
\caption[]{Relative error from quenching in $f_B$, $f_{B_s}$,
$B_B$ and $B_{B_s}$ at one-loop in QChPT.  For explanation and
assumptions, see text.  From \cite{Sh96}.}
\end{figure}

There is an extensive body of work on QChPT, and I do not
have space here to properly review even a 
substantial fraction of it.
There has been work on the inclusion of other hadrons (baryons:
\cite{La96}, vector and tensor mesons: \cite{Bo97,Ch97},
heavy baryons: \cite{Chi97}), weak matrix elements
($B_K$: \cite{Sh92}, $f_B$ and $B_B$: \cite{Bo95,Sh96},
$K^+\to\pi^+\pi^0$ and $B_K$: \cite{Go97}, baryon axial
charge: \cite{Ki96}), pion scattering (\cite{Be96})
and formalism (\cite{Co97}).  For work on fitting lattice
data to results from QChPT, I refer to the Proceedings of
the Lattice 96 and Lattice 97 conferences, in particular the
reviews by \cite{GoS97}, \cite{Sh971} and \cite{Ok97}, and
references therein.

Here, let me show a sample result of a case where the GBs
are coupled to other hadrons, in this case the $B$ meson.
Fig. 2, taken from \cite{Sh96}, shows the relative error
from quenching for $f_B$, $f_{B_s}$, $B_B$ and $B_{B_s}$,
as a function of the many parameters that are not determined
by ChPT, from a one-loop ChPT calculation.  
Note that more parameters typically show up than
in the case of pure GB physics, because of the various
coupling constants between the GBs and the heavy sector.
$g$ ($g_Q$) is the $B$-pion coupling in the unquenched
(quenched) theory, $g'$ is the $B$-$\eta'$ coupling (which
is suppressed in $1/N$).  The figure assumes $g_Q=g$, and
values as shown (the value for $g$ is from an estimate in
the unquenched theory).  All $O(p^4)$ LECs have arbitrarily
been set to zero, while the cutoff is chosen at $\Lambda=1$~GeV.
Fig. 2 shows that, under these assumptions, quenching errors
in these quantities can be large.  As such, it gives us
useful information, but it should be stressed that this is
{\em not} a calculation of quenching errors, because of the
many possibly unjustified choices made for all the parameters.
This situation is typical in the case that we consider the
effective theory for GBs coupled to other hadrons.

\section{Pion Scattering}
LQCD computations give us access to euclidean correlation
functions only, and, lacking the possibility to analytically
continue them to Minkowski space, it appears impossible to
extract information on scattering lengths from the lattice.
However, we can make use of the fact that lattice computations
are necessarily done in a finite volume, and use this as a 
probe of the interactions between two pions.  More precisely,
one can compute the energy of a state with two pions at rest
from the euclidean correlation function (for $I=2$, a similar
correlation function can be defined for $I=0$)
\begin{eqnarray}
C_{I=2}&=&\langle 0|\pi^+(t)\pi^+(t)\pi^-(0)\pi^-(0)|0\rangle
\label{corr} \\
&&=\sum_{|\alpha\rangle}{\rm e}^{-E_\alpha t}
|\langle\alpha|\pi^-(0)\pi^-(0)|0\rangle|^2 \nonumber \\
&&=Z{\rm e}^{-2m_\pi t}\left(1-\Delta E\;t+O(t^2)\right)+
{\rm excited\ states}, \nonumber
\end{eqnarray}
where $\pi^-(0)$ creates a zero-momentum $\pi^+$ at time $0$,
and $\pi^+(t)$ annihilates a zero-momentum $\pi^+$ at time $t$.
We inserted a complete set of states $\{|\alpha\rangle\}$,
and assume that we can take $t$ large enough to project out
the lowest-energy state.  This is the state with two pions
at rest in finite volume, with $E_{2\pi}=2m_\pi+\Delta E$,
where $\Delta E$ is the energy shift due to finite volume
effects (in infinite volume, two pions at rest do not
interact).  We expanded in $\Delta E\;t$, because it is in
this form that the energy shift will show up in a ChPT
calculation.  Also, it was this form which has been fitted
to lattice data by various groups in attempts to determine
$\Delta E$ (\cite{It90,Gu92,Ku93}).  From (\ref{corr}) we
see that $\Delta E\;t$ has to be small enough for the
expansion on the last line to be valid (but of course
$t$ has to be large enough to project out the excited
states).

As mentioned above, $\Delta E$ is a measure of the interactions
between the two pions, due to their confinement to a finite
volume.  A precise connection between $\Delta E$ and the
pion scattering length was given by \cite{Lu86}:
\begin{equation}
\Delta E=-\frac{4\pi a_0}{m_\pi L^3}
\left(1-2.837297\frac{a_0}{L}+6.375183
\frac{a_0^2}{L^2}\right)+O\left(\frac{1}{L^6}\right),
\label{energy}
\end{equation}
where $a_0$ is the infinite volume scattering length
for the corresponding channel ($I=0$ or $2$).  While
the proof of (\ref{energy}) is more general, the r.h.s.
can be recovered term by term at tree level, one loop,
two loops, etc. in ChPT. It is instructive to look at this
in some detail, because it will tell us how things change
in the QA.  

Consider the one-loop diagram in which the pions are
created at time $0$, scatter at time $t_2>0$, rescatter at
time $t_1>t_2$, and then are annihilated at time $t>t_1$.
(There are many other one-loop contributions of course,
but it is this one that leads to the $1/L^4$ term in 
(\ref{energy}).)
Since the initial and final pions have zero spatial momentum,
their propagators are of the form ${\rm exp}(-m_\pi t)$ if
they ``travel" a time $t$; the expression for this diagram
looks like
\begin{equation}
\lambda^2L^3{\rm e}^{-2m_\pi t}\;
\frac{1}{L^3}\sum_{\vec k}{\rm e}^{-2(E({\vec k})-m_\pi)
(t_1-t_2)}, 
\ \ \ \ E({\vec k})=\sqrt{m_\pi^2+{\vec k}^2}. \label{diagram}
\end{equation}
The second exponential takes into account that between the
two scatterings, the two pions can have arbitrary momenta
$\vec k$ and $-{\vec k}$.  The overall factor $L^3$ comes
from integrating the diagram over the spatial volume, and
$\lambda\propto m_\pi^2/(4\pi f_\pi)^2$ is the interaction
strength.  This expression has to be integrated over $t_1$
and $t_2$, where here we are interested in the contribution
with $0<t_2<t_1<t$.  This integration leads to
\begin{equation}
t\;\frac{1}{L^3}\sum_{{\vec k}\ne 0}\frac{1}{E({\vec k})-m_\pi}
\label{denom}
\end{equation}
(the ${{\vec k}=0}$ term contributes to the $O(t^2)$ term in
(\ref{corr})).  In infinite volume, we may replace
$\frac{1}{L^3}\sum_{{\vec k}\ne 0}$ by $\int d^3k$, and
(after renormalization) we get a contribution $\sim\lambda^2/
L^3$ to $\Delta E$. (The $L^3$ in (\ref{diagram})
turns into a $1/L^3$ after dividing by the factor $L^6$ which
normalizes the contribution $L^6{\rm exp}(-2m_\pi t)$ of two
noninteracting pions.)  This just gives us a one-loop
correction to $a_0$ in the $1/L^3$ term in (\ref{energy}).
In a finite volume, the smallest
momenta $k\sim 1/L$ cause the sum to deviate from the integral
most prominently.  Expanding $E({\vec k})-m_\pi\sim
1/L^2$ (keeping track of the $L$ dependence only), 
the sum (\ref{denom}) gives a 
contribution $\sim t/L$.  This leads to an additional
finite volume contribution to $\Delta E$ of order
$\lambda^2/L^4\sim a_0^2/L^4$, cf. the second term
in (\ref{energy}). 

We can now also see how quenching, and in particular, the
special role of the $\eta'$, will change this result.  In the
$I=0$ channel, the intermediate mesons in the diagram
I just discussed can also be singlets, and the two-point
vertex $-(\mu^2+\alpha p^2)$ can appear on these internal lines.
Let us consider the case of one such insertion at time $t_X$.  
The expression
for the diagram is similar to (\ref{diagram}), but now we have
to integrate also over $t_X$!  Considering the 
contribution where $t_2<t_X<t_1$, we pick up an extra factor
$t_1-t_2$ after doing the integral over $t_X$, and from the
integration over $t_1$ and $t_2$ we now get
\begin{equation}
t\;\frac{1}{L^3}\sum_{{\vec k}\ne 0}
\frac{1}{(E({\vec k})-m_\pi)^2}\;.
\label{denomsq}
\end{equation}
Going through the same argument as before, this leads to an extra
factor $L^2$ in the finite-volume energy shift, i.e. a 
contribution $\sim\delta/L^2$ to $\Delta E$, quite unlike the
prediction of L\"uscher's formula!  Even worse, if we insert
the $\eta'$ two-point vertex on {\em both} internal lines,
we find another factor $L^2$ enhancement, and a one-loop
contribution to $\Delta E$ of order $\delta^2/L^0=\delta^2$.  
These ``enhanced finite volume" corrections are another example
of the serious infrared problems that affect the quenched
theory, due to the double pole in (\ref{etapropfull}).  For
a complete analysis (and real calculation), see
\cite{Be96}.  We just note here that also the rest of
the structure of (\ref{energy}) is not reproduced in the
quenched theory.  For $I=2$ the problem is less serious,
because no enhanced corrections occur (the internal lines
cannot be singlets in this case).

\begin{figure}[t]
\vspace{-1.0cm}
\includegraphics{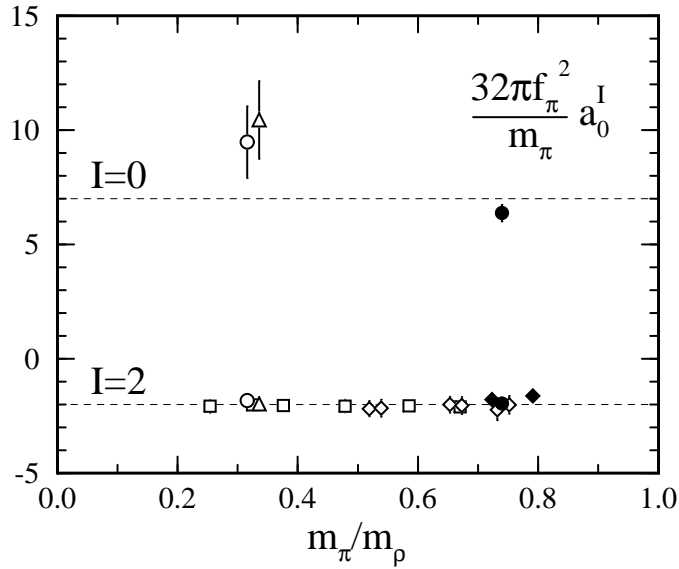}
\vspace{9.5cm}
\caption[]{Pion scattering lengths 
as a function of $m_\pi/m_\rho$.
Errors are statistical only.  The squares and diamonds
are from \cite{Gu92}, all other points are from \cite{Ku93},
from which this figure was taken.}
\end{figure}

We conclude that QChPT does not only break down for small
$L$, but also for large $L$, since the tree-level term in
$\Delta E$ goes like $1/L^3$, while there are one-loop
corrections which go like a lower power of $1/L$.  In fact,
one has to worry whether QChPT applies for any choice
of parameter values in this case! It is
therefore very interesting to compare these results with
quenched numerical computations of the pion scattering 
lengths. 
 
In Fig. 3, which I took from \cite{Ku93}, lattice results
for $a_0^{I=0,2}$ are shown, in units of tree-level ChPT
predictions, as a function of $m_\pi/m_\rho$.  Also results
from \cite{Gu92} for $I=2$ are included (squares and diamonds).
The volume is $12^3\times 20$ in lattice units, $\beta=5.7$,
corresponding to $a\approx 0.2$~fm.
We see that, within errors (only statistical errors are shown),
the lattice results agree with ChPT at tree level.  In
\cite{Be96} we considered the size of the one-loop QChPT 
corrections for the parameters of these lattice computations,
and found that they are always smaller than 20\% of the 
tree-level values.  However, we cannot directly compare
the lattice results with ChPT because of the presence of 
other systematic errors, most notably scaling violations
($a$ is too big), and contamination of excited states
($t$ in (\ref{corr}) may not have been large enough to reliably
project out the lowest state, see \cite{Be96}).  

\section{$K^+\to\pi^+\pi^0$ Decay}
LQCD practitioners have had a long-standing interest in
computing nonleptonic $K$-decay matrix elements, because of
such prominent experimental observations as the $\Delta I=
1/2$ rule, and their importance in determining Standard
Model parameters (the CKM-matrix).  I will restrict myself
here to $K^+\to\pi^+\pi^0$.  References \cite{Ga88,Be89}
report on the status of past lattice computations of  
$\langle\pi^+\pi^0|O_4|K^+\rangle$, while \cite{Is97}
reports on a very recent computation. Here
$O_4=({\overline s}_L\gamma_\mu d_L)
({\overline u}_L\gamma^\mu u_L)+({\overline s}_L\gamma_\mu u_L)
({\overline u}_L\gamma^\mu d_L)-({\overline s}_L\gamma_\mu d_L)
({\overline d}_L\gamma^\mu d_L)$ is the $\Delta I=3/2$, $\Delta S
=1$ component of the SU(3)$_L$ 27-plet responsible for this decay.
In ChPT, we have, with $\Sigma$ the nonlinear GB-field,
\begin{equation}
O_4=\alpha_{\scriptscriptstyle 27}t^{ij}_{kl}
(\Sigma\partial_\mu\Sigma^\dagger)_i^{\ k}
(\Sigma\partial^\mu\Sigma^\dagger)_j^{\ l}
+O(p^4)\ {\rm operators}, \label{ofour}
\end{equation}
where $t^{ij}_{kl}$ projects out the appropriate component,
$\alpha_{\scriptscriptstyle 27}$ 
is a free parameter, and the $O(p^4)$ operators
were listed in \cite{Ka90}.
In \cite{Go97}, $\langle\pi^+\pi^0|O_4|K^+\rangle$ was 
calculated to one loop in order to compare the real world
with lattice results.  Since the $O(p^4)$ coefficients
are not or very
poorly known, we set them equal to zero, and estimated
the error from this by considering two values of the cutoff
$\Lambda=770$~MeV$/1$~GeV.  The real-world value then is
(for tree level, see \cite{Do82}, for one loop with $m_\pi
=0$, see \cite{Bij84}) 
\begin{equation}
\langle\pi^+\pi^0|O_4|K^+\rangle=
{{12i \alpha_{\scriptscriptstyle 27}}
\over{\sqrt{2}f_\pi^3}}
\left(m^2_K -m^2_\pi \right)\times
\left( 1+
{{0.63,\ \Lambda=1\phantom{70}\ {\rm GeV}}\atop
{0.36,\ \Lambda=770\ {\rm MeV}}
}\right).\label{rwdecay}
\end{equation}

On the lattice, one computes the matrix element from
\begin{eqnarray}
C(t_2,t_1)&=&\langle 0|\pi^+(t_2)\pi^0(t_2)\;O_4(t_1)
\;K^-(0)|0\rangle \label{corrk} \\
&{\buildrel{\scriptstyle {t_2 \gg t_1 \gg 0}} \over
\longrightarrow}&{\rm e}^{-E_{2\pi} (t_2-t_1)} {\rm e}^{-m_K t_1}
\nonumber \\ 
&&\times
{{\langle 0|\pi^+(0)\pi^0(0)|\pi^+\pi^0\rangle\langle\pi^+\pi^0
|O_4(0)|K^+\rangle\langle K^+|K^-(0)|0\rangle}
\over
{\langle\pi^+\pi^0|\pi^+\pi^0\rangle\langle K^+|K^+\rangle}
}, \nonumber 
\end{eqnarray}
where all mesons are taken to be at rest.  In actual
lattice computations, degenerate quark masses were used, so
that $m_K=m_\pi$.  The matrix element in (\ref{corrk}) therefore
is not the one we want: it is unphysical because of the choice
of masses and external momenta.  Furthermore, there are
power-like finite volume effects, and the QA has been used in
all computations. All these systematic effects can be estimated
in one-loop ChPT (the unphysical choice of masses and momenta
already at tree level in ChPT (\cite{Be89}).

To one loop in ChPT the physical matrix element
and the unphysical quenched lattice matrix element 
(after extrapolation to the continuum
limit) are related by (\cite{Go97})
\begin{equation}
{\langle\pi^+\pi^0|O_4(0)|K^+\rangle
}_{phys}=Y\;{\alpha_{\scriptscriptstyle 27}\over
\alpha^q_{\scriptscriptstyle 27}}\left({f_q\over f}\right)^3
{{m_K^2-m_\pi^2}\over 2M^2_\pi}\;
{\langle\pi^+\pi^0|O_4(0)|K^+\rangle
}^{quenched}_{unphys},\label{qrelat}
\end{equation}
with
\begin{eqnarray}
Y&=&{
{1+{{\phantom{-}0.089,\ \Lambda =1\phantom{70}\ {\rm GeV}}\atop{-0.015,\
\Lambda=770\ {\rm MeV}} }}
\over
{1+{M^2_\pi\over(4\pi
F_\pi)^2}\left[-3\log{M^2_\pi\over\Lambda^2_q}+F(M_\pi
L)\right]}}\ , \label{Y} \\
F(m_\pi L)&=&{17.827/{(m_\pi L)}}+{12\pi^2 /{(m_\pi L)}^3}.\label{F}
\end{eqnarray}
Here parameters with a subscript $q$ are those of the quenched theory,
and $M_\pi=M_K$, $F_\pi$ refer to the lattice quantities. 
At tree level, we would have $Y=1$. $O(p^4)$
LECs have been set to zero. Some remarks are in order:
\begin{itemize}
\item For degenerate quark masses, $O_4$ does not couple
to the $\eta'$, which explains why (\ref{Y}) does not depend on
$\delta$ and $\alpha$.  For nondegenerate quark masses, such 
dependence would show up, presumably making the QA less interesting
in that case.
\item The ratios $\alpha_{\scriptscriptstyle 27}/
\alpha_{\scriptscriptstyle 27}^q$ and $f_q/f$ are not known, and,
in the following discussion of lattice results, we will arbitrarily
set them equal to one. 
\item The $O(p^4)$ LECs are basically ``absorbed" into
the cutoff $\Lambda$.  As before, 
we will take the cutoff equal to 770~MeV or 1~GeV {\em independently}
in the quenched and unquenched theories, and take the spread as an
indication of the systematic error associated with our lack of
knowledge of these LECs.
\end{itemize}
 
\begin{figure}[t]
\vspace{-4.0cm}
\epsfxsize=1.0 \hsize
\epsfbox{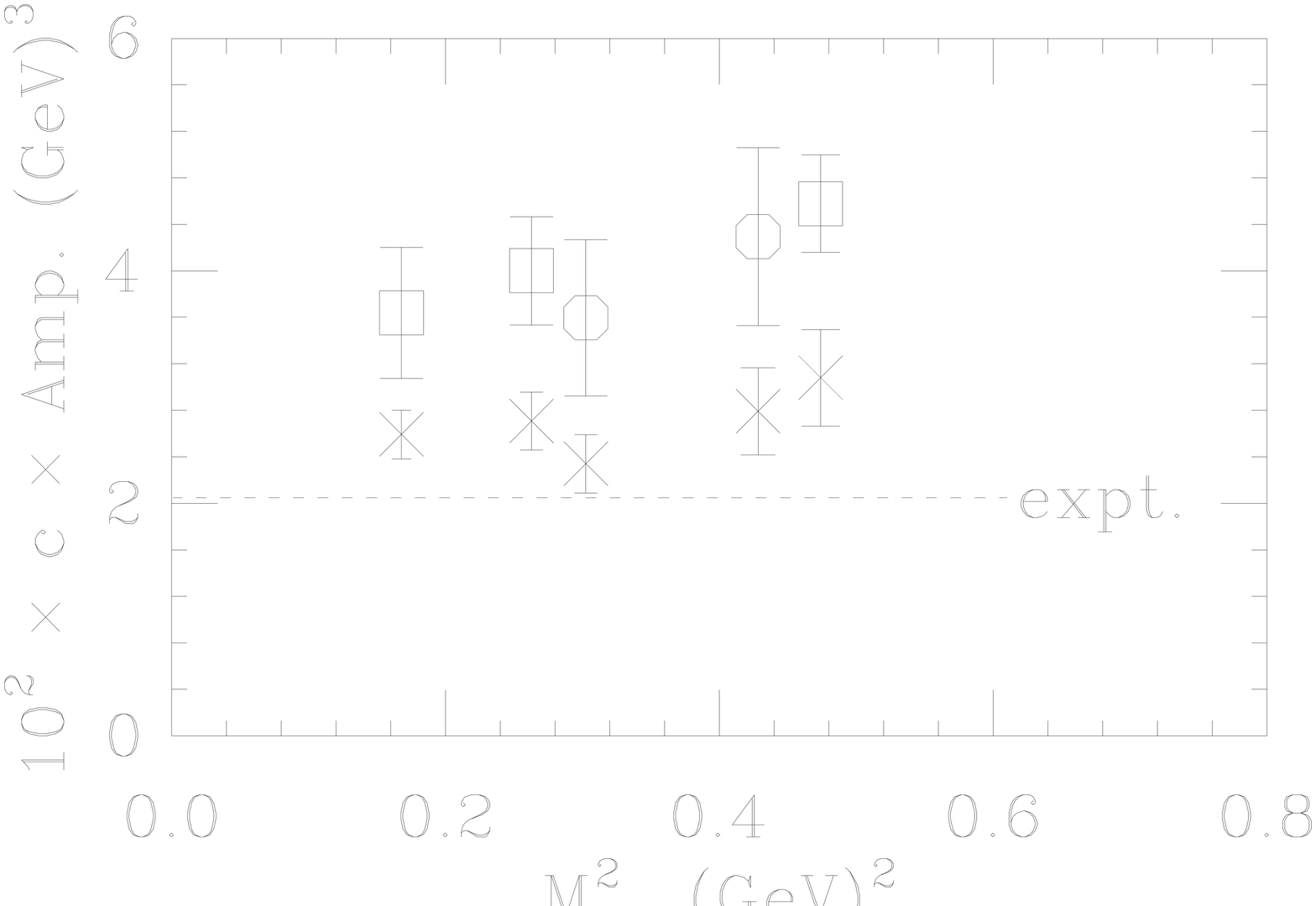}
\vspace{1cm}
\caption[]{$K^+\to\pi^+\pi^0$ decay amplitude as a function of
pion mass. Open symbols: data from \cite{Be89} (squares:
$16^3\times 25$ (or $\times 33$), $\beta=5.7$; octagons: 
$24^3\times 40$, $\beta=6$); crosses: including the correction
factor $Y$. The constant 
$c=2\sqrt{2}/(G_F\sin\theta_c\cos\theta_c)$.}
\end{figure}

In Fig. 4 I show lattice data for the matrix element from
\cite{Be89} (open symbols), where the tree-level correction
factor $(m_K^2-m_\pi^2)/2M_\pi^2$ in (\ref{qrelat}) was already
taken into account.  The errors on these points are 
statistical only, and we left out points at pion masses
with $M_\pi>m_\rho$ and/or at smaller volumes.  
The crosses show what we obtain if we
``correct" the central values of 
each of these points with the factor $Y$ 
(evaluated at the appropriate pion mass and volume).
The errors on these points indicate the spread from choosing
different combinations of $\Lambda$ and $\Lambda_q$.
We see that at all points $Y<1$, and that therefore the 
one-loop corrections reduce the discrepancy between lattice
data and the experimental result.  However, since one-loop
effects are rather substantial, two-loop effects can probably
not be neglected.  Also, again, scaling violations and various
other systematic effects are not taken into account.  For
a more complete discussion, see \cite{Go97}.

\begin{figure}[t]
\epsfxsize=1.0 \hsize
\epsfbox{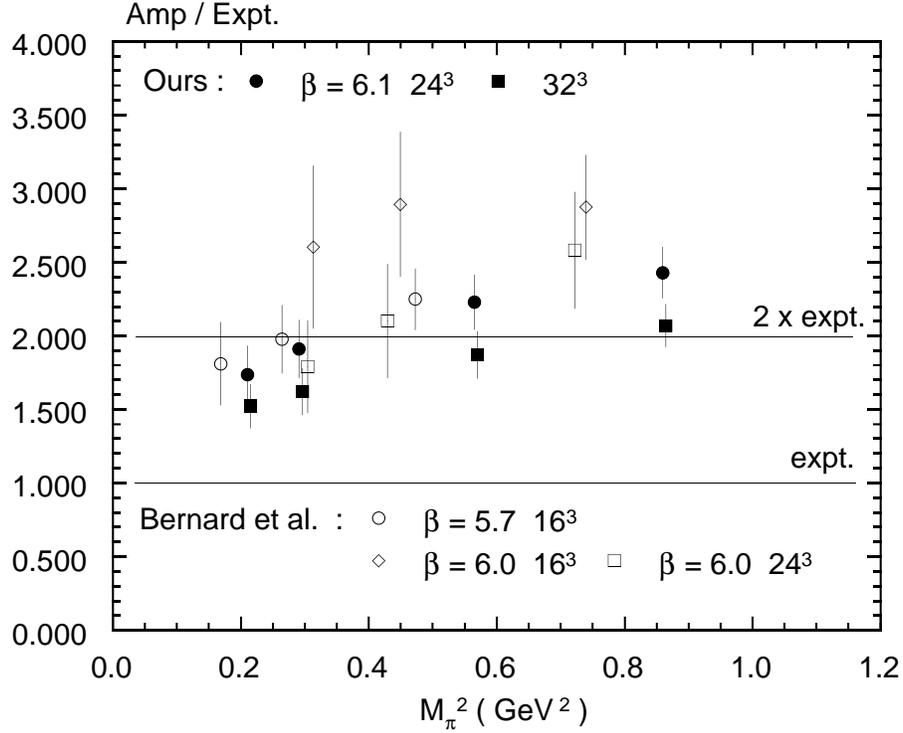}
\caption[]{$K^+\to\pi^+\pi^0$ decay amplitude as a function of
pion mass, in units of the experimental value. Figure provided
by JLQCD (\cite{Is97}).  Also shown are the data
from \cite{Be89}.  Tree-level corrected.}
\end{figure}

Very recently, the computation was done again by the JLQCD
collaboration, with larger volumes, and larger $\beta$
(\cite{Is97}).
The results are shown in Fig. 5 (without the one-loop
correction factor $Y$, along with the data of \cite{Be89})
and Fig. 6 (with the factor $Y$ taken into account).  If the
unquenched cutoff $\Lambda=\Lambda^{\rm cont}=1$~GeV is 
chosen, the values in Fig. 6 would come out about 10\%
higher.  Again, I should stress that, because of the 
uncertainties in the estimates for $Y$, the lack of
knowledge of the ratios $\alpha_{\scriptscriptstyle 27}/
\alpha_{\scriptscriptstyle 27}^q$ and $f_q/f$, and various
systematic effects which cannot be estimated in ChPT, one
can only conclude that one-loop ChPT reduces the discrepancy
between lattice and experiment.
 
\begin{figure}[t]
\epsfxsize=1.0 \hsize
\epsfbox{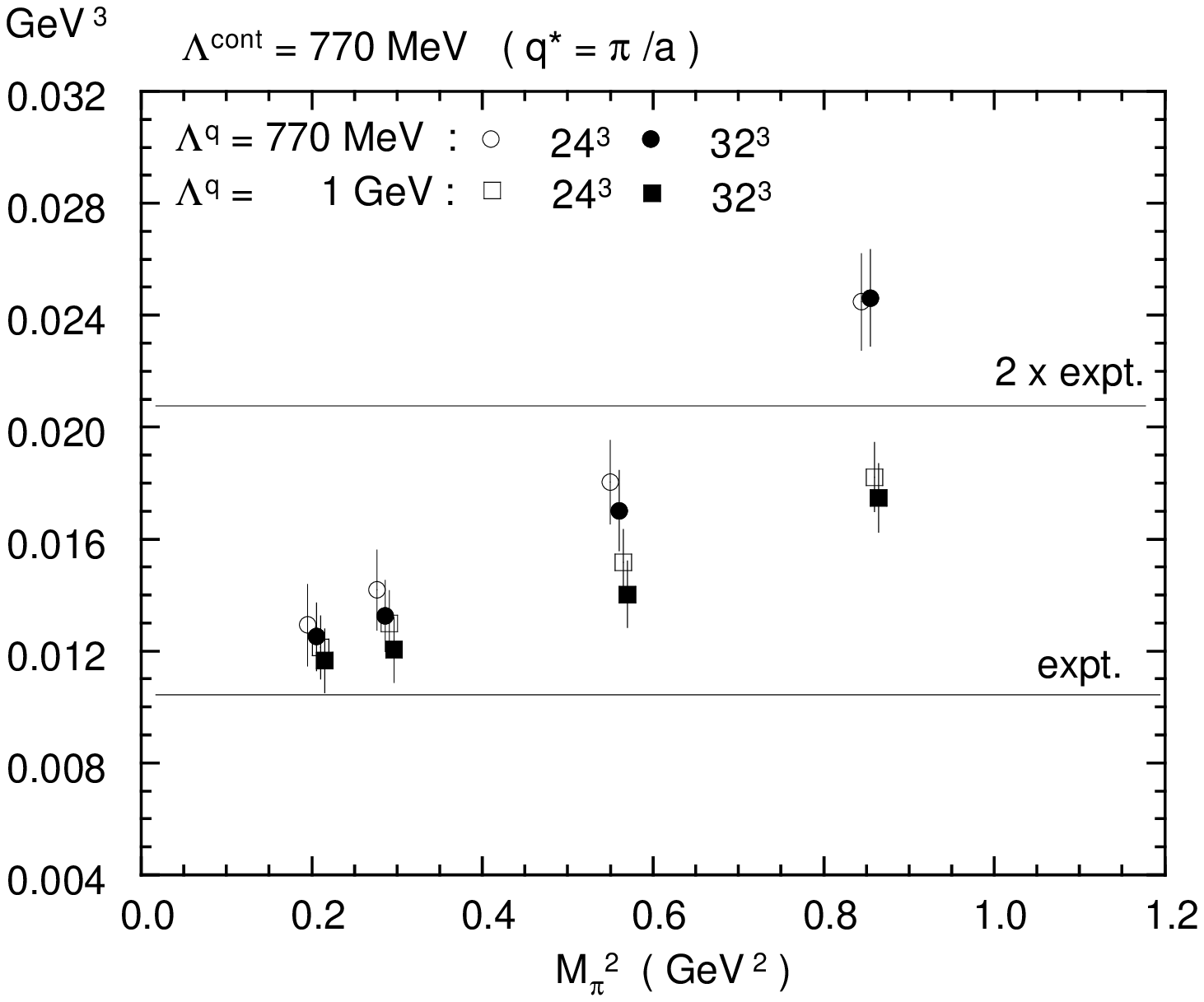}
\caption[]{$K^+\to\pi^+\pi^0$ decay amplitude as a function of
pion mass, corrected with the one-loop factor $Y$. 
Figure provided by JLQCD (\cite{Is97}).}
\end{figure}

\section{Conclusion}
It is clear that one-loop ChPT plays an important role in
understanding current LQCD results.  However, numerical
computations are not yet at a level of precision that 
$O(p^4)$ LECs can be reliably extracted from the lattice.
The most extensive (small quark masses, large volumes)
computations are done in the (partially) QA, to which
ChPT can be adapted systematically.  The fundamental
difference between QCD and its quenched relative shows up
in full force at one-loop in ChPT: the nonanalytic terms
in the QA are in general very different from those of the
full theory.  The safest way to look at these one-loop
differences is to take them as an indication of the 
systematic error made by using the QA.

We have seen that the quenched theory is afflicted with
infrared divergences which show up as a diverging chiral
limit, and a diverging infinite volume limit.  This is all
due to the special role of the $\eta'$ in the QA.  It is
therefore important to test the predictions of QChPT 
against lattice computations, as long as we will be using
the (partially) QA.

\subsubsection{Acknowledgements.} I would like to thank the organizers
and participants of the Workshop for creating a stimulating atmosphere.
I would also like to thank Claude Bernard, Steve Gottlieb,
Naruhito Ishizuka, Ka Chun Leung, Steve Sharpe
and Akira Ukawa for many discussions on various aspects of the work
described here and for providing me with the figures,
and Pierre van Baal and the University of Leiden for hospitality
while preparing this talk.  This work
was supported in part by the US Department of Energy through an
Outstanding Junior Investigator grant.
 
%

%
%


\begin{thebibliography}
%
\bibitem{}{BeGo92}{Bernard and Golterman (1992)}
C.~Bernard and M.~Golterman, Phys. Rev. {\bf D46} (1992) 853.
%
\bibitem{}{BeGo93}{Bernard and Golterman (1993)}
C.~Bernard and M.~Golterman,
Nucl. Phys. {\bf B} (Proc. Suppl.) {\bf 30} (1993) 217. 
%
\bibitem{}{BeGo94}{Bernard and Golterman (1994)}
C.~Bernard and M.~Golterman, Phys. Rev. {\bf D49} (1994) 486.
%
\bibitem{}{Be96}{Bernard and Golterman (1996)}
C.~Bernard and M.~Golterman, Phys. Rev. {\bf D53} (1996) 476.
%
\bibitem{}{Be89}{Bernard and Soni (1989)}
C.~Bernard and A.~Soni, Nucl. Phys. {\bf B} (Proc. Suppl.)
{\bf 9} (1989) 155.
%
\bibitem{}{Ba96}{Bhattacharya and Gupta (1996)}
T.~Bhattacharya and R.~Gupta, Phys. Rev. {\bf D54} (1996) 1155.
%
\bibitem{}{Bij84}{Bijnens et al. (1984)}
J.~Bijnens, H.~Sonoda and M.~Wise, Phys. Rev. Lett. {\bf 53}
(1984) 2367.
%
\bibitem{}{Bo95}{Booth et al. (1995)}
M.~Booth, Phys. Rev. {\bf D51} (1995) 2338; hep-ph/9412228.
%
\bibitem{}{Bo97}{Booth et al. (1997)}
M.~Booth et al., Phys. Rev. {\bf D55} (1997) 3092.
%
\bibitem{}{Chi97}{Chiladze (1997)}
G.~Chiladze, hep-ph/9704426.
%
\bibitem{}{Ch97}{Chow and Rey (1997)}
C.-K.~Chow and S.-J.~Rey, hep-ph/9708432.
%
\bibitem{}{Co97}{Colangelo and Pallante (1997)}
G.~Colangelo and E.~Pallante, hep-lat/9702019; hep-lat/9708005.
%
\bibitem{}{DeW}{DeWitt (1984)}
B.~DeWitt, {\it Supermanifolds}, Cambridge University Press (1984).
%
\bibitem{}{Do82}{Donoghue et al. (1982)}
J.~Donoghue, E.~Golowich and B.~Holstein, Phys. Lett. {\bf B119}
(1982) 412.
%
\bibitem{}{Ga97}{Gasser (1997)}
J.~Gasser, these proceedings.
%
\bibitem{}{GaLe85}{Gasser and Leutwyler (1985)} 
J.~Gasser and H.~Leutwyler, Nucl. Phys. {\bf B250}, 465 (1985).
%
\bibitem{}{Ga87}{Gasser and Leutwyler (1987)}
J.~Gasser and H.~Leutwyler, Phys. Lett, {\bf B184} (1987) 83.
%
\bibitem{}{Ga88}{Gavela et al. (1988)}
M.B.~Gavela et al., Nucl. Phys. {\bf B306} (1988) 677.
%
\bibitem{}{Go94}{Golterman (1994)}
M.~Golterman, Acta Phys. Pol. {\bf 25} (1994) 1731.
%
\bibitem{}{Go97}{Golterman and Leung (1997)}
M.~Golterman and K.-C.~Leung, Phys. Rev. {\bf D56} (1997) 2950.
%
\bibitem{}{GoS97}{Gottlieb (1997)}
S.~Gottlieb, Nucl. Phys. {\bf B} (Proc. Suppl.) {\bf 53} (1997) 155.
%
\bibitem{}{It90}{Guagnelli et al. (1990)}
M.~Guagnelli, E.~Marinari and G.~Parisi, Phys. Lett. {\bf B240} (1990)
188.
%
\bibitem{}{Gu92}{Gupta et al. (1992)}
R.~Gupta et al., Nucl. Phys. {\bf B383} (1992) 309;
Phys. Rev. {\bf D48} (1993) 388.
%
\bibitem{}{Pa81}{Parisi et al. (1981)}
H.~Hamber and G.~Parisi, Phys. Rev. Lett. {\bf 47} (1981) 1792;
E.~Marinari, G.~Parisi and C.~Rebbi, Phys. Rev. Lett. {\bf 47} (1981)
1795.
%
\bibitem{}{Is97}{Ishizuka et al. (1997)}
N.~Ishizuka, for JLQCD, to be published in the Proceedings of the
XVth International Conference on Lattice Field Theory,
Edinburgh, Scotland (1997).
%
\bibitem{}{Ka90}{Kambor et al. (1990)}
J.~Kambor, J.~Missimer and D.~Wyler, Nucl. Phys. {\bf B346} (1990)
17.
%
\bibitem{}{Ki96}{Kim and Kim (1996)}
M.~Kim and S.~Kim, hep-lat/9608091.
%
\bibitem{}{Ku93}{Kuramashi et al. (1993)}
Y.~Kuramashi et al., Phys. Rev. Lett. {\bf 71} (1993) 2387;
Phys. Rev. {\bf D52} (1995) 3003.
%
\bibitem{}{La96}{Labrenz and Sharpe (1996)}
J.~Labrenz and S.~Sharpe, Phys. Rev. {\bf D54} (1996) 4595.
%
\bibitem{}{Lu86}{L\"uscher (1986)}
M.~L\"uscher, Comm. Math. Phys. {\bf 105} (1986) 153;
Nucl. Phys. {\bf B354} (1991) 531.
%
\bibitem{}{Mo87}{Morel (1987)}
A.~Morel, J. Physique {\bf 48} (1987) 111.
%
\bibitem{}{Sh90}{Sharpe (1990)}
S.~Sharpe, Phys. Rev. {\bf D41} (1990) 3223; S.~Sharpe, 
Nucl. Phys. {\bf B} (Proc. Suppl.) {\bf 17} (1990) 146;
G.~Kilcup et al., Phys. Rev. Lett. {\bf 64} (1990) 25. 
%
\bibitem{}{Sh92}{Sharpe (1992)}
S.~Sharpe, Phys. Rev. {\bf D46} (1992) 3146.  
%
\bibitem{}{Sh971}{Sharpe (1997a)}
S.~Sharpe, Nucl. Phys. {\bf B} (Proc. Suppl.) {\bf 53} (1997) 181.
%
\bibitem{}{Sh972}{Sharpe (1997b)}
S.~Sharpe, hep-lat/9707018.
%
\bibitem{}{Sh96}{Sharpe and Zhang (1996)}
S.~Sharpe and Y.~Zhang, Phys. Rev. {\bf D53} (1996) 5125.
%
\bibitem{}{Ve79}{Veneziano (1979)}
G.~Veneziano, Nucl. Phys. {\bf B159} (1979) 213.
%
\bibitem{}{WeS79}{Weinberg (1979)}
S.~Weinberg, Physica {\bf 96A}, 327 (1979).
%
\bibitem{}{WeD82}{Weingarten (1982)}
D.~Weingarten, Phys. Lett. {\bf B109} (1982) 57.
%
\bibitem{}{Wi79}{Witten (1979)}
E.~Witten, Nucl. Phys. {\bf B159} (1979) 269.
%
\bibitem{}{Ok97}{Yoshie (1997)}
T.~Yoshie, to be published in the Proceedings of the
XVth International Conference on Lattice Field Theory,
Edinburgh, Scotland (1997) (hep-lat/9711017).
%
%
\end{thebibliography}
\end{document}